\documentclass[twocolumn,aps,prl,superscriptaddress,amsmath,amssym,floatfix,longbibliography,notitlepage,amsbook]{revtex4}
\usepackage[colorlinks=true,linkcolor=blue,citecolor=blue,filecolor=green,urlcolor=blue]{hyperref}
\usepackage{graphicx}
\usepackage{algorithm}
\usepackage{algpseudocode}
\usepackage{physics}

\def\be{\begin{equation}}
\def\ee{\end{equation}}

\begin{document}

\newcommand{\indep}{\perp \!\!\! \perp}

\title{Reconstructing Superoscillations Buried Deeply in Noise}

\author{Derek D. White}
\affiliation{Institute for Quantum Studies, Chapman University, Orange, CA 92866, USA}

\author{Shunxing Zhang}
\affiliation{Institute for Quantum Studies, Chapman University, Orange, CA 92866, USA}

\author{Barbara {\v S}oda}
\affiliation{Department of Applied Mathematics and Department of Physics, University of Waterloo, and Perimeter Institute for Theoretical Physics, Waterloo, Ontario, Canada}

\author{Achim Kempf}
\affiliation{Department of Applied Mathematics and Department of Physics, University of Waterloo, and Perimeter Institute for Theoretical Physics, Waterloo, Ontario, Canada}

\author{Daniele C. Struppa}
\affiliation{The Donald Bren Presidential Chair in Mathematics, Chapman University, Orange, CA, 92866, USA}

\author{Andrew N. Jordan}
\affiliation{Institute for Quantum Studies, Chapman University, Orange, CA 92866, USA}
\affiliation{The Kennedy Chair in Physics, Chapman University, Orange, CA 92866, USA}
\affiliation{Center for Coherence and Quantum Optics, University of Rochester, Rochester, NY 14627, USA}
\affiliation{Department of Physics and Astronomy, University of Rochester, Rochester, NY 14627, USA}

\author{John C. Howell}
\affiliation{Institute for Quantum Studies, Chapman University, Orange, CA 92866, USA}
\affiliation{Racah Institute of Physics, The Hebrew University of Jerusalem, Jerusalem, Israel, 91904}
\email{johhowell@chapman.edu}

\begin{abstract}
We utilize a method using frequency combs to construct waves that feature superoscillations - local regions of the wave that exhibit a change in phase that the bandlimits of the wave should not otherwise allow. This method has been shown to create superoscillating regions that mimic any analytic function - even ones well outside the bandlimits - to an arbitrary degree of accuracy. We experimentally demonstrate that these waves are extremely robust against noise, allowing for accurate reconstruction of a superoscillating target function thoroughly buried in noise. We additionally show that such a construction can be easily used to range-resolve a signal well below the commonly accepted fundamental limit.
\end{abstract}

\maketitle
\section{Introduction}
Superoscillations \cite{Aharonov-preprint,popescu1991multi,berry2006evolution,kempf2018four,berry2019roadmap} are temporally local frequencies in a signal that are outside the signal's bandlimits. This phenomenon seemingly defies our typical notions of Fourier limits. These faster-than-expected oscillations are closely related to the physics of weak quantum measurements \cite{Aharonov-preprint,Aharonov2011Aug,dressel2014colloquium,jordan2022super,jordan2024quantum}. 
The ability to generate fast responses or high angular resolution much faster than a signal's maximum frequency opens up a trove of possible applications. To date, superoscillations have been used in applications such as microscopy \cite{huang2009super,rogers2012super,Kozawa2018Feb,gbur2019using, Wong2013}, holography \cite{hao2017three}, ultra-fast optics \cite{eliezer2017breaking}, and beam focusing \cite{makris2011superoscillatory,greenfield2013experimental,yuan2017achromatic}, among others \cite{wong2011b, berry2019roadmap, Wong2017, Yuan2014, Xiao2024}. Berry, a pioneer in the field, describes superoscillations as being "able to play Beethoven's Ninth Symphony with a 1 Hz bandlimited signal" \cite{berry1994faster}.

An exciting yet minimally explored application of superoscillations lies in the field of radar \cite{Wong2011, wong2012superoscillatory}, where the ability to range-resolve more than one object is limited by the bandwidth ($B$) of the signal. The traditionally accepted minimum resolvable separation distance ($\Delta R$) between scatterers is given by
\begin{equation}
    \Delta R = \frac{c}{2 B},
\label{range_resolved}
\end{equation}
where $c$ is the speed of light \cite{Wehner_1995, Nuss2020, Sengupta2024, Lang2020, Vasconcelos2023}. When two scattering objects are separated by less than half the inverse bandwidth of the signal, the reflections overlap, appearing as a single pulse. In such cases, while there may be noticeable changes in the return signal compared to its outbound counterpart, it lacks sufficient distortion to identify the presence of two distinct scatterers. By utilizing a superoscillating signal's ability to locally increase its effective bandwidth beyond its global limits, it should therefore be possible to enhance range resolution proportionally to the local bandwidth gain.

Superoscillatory waves come with a cost, however \cite{berry2019roadmap,EnergyExpense,Berry_2017}. This is reflected in the "Roadmap on Superoscillations" \cite{berry2019roadmap}, in which a theme is repeated wherein the wave's amplitude in the finite region where superoscillations exist decreases exponentially relative to the "lobes" outside the region of interest as the superoscillating region's frequency increases. These huge disparities between the region of interest and the lobes makes the superoscillations "vulnerable to noise" or "highly sensitive to noise," resulting in "rapid disappearance of superoscillations." Thus, superoscillations often require "exorbitantly high signal-to-noise ratios." 

Frequency combs (waves constructed from a finite series of discrete frequencies) have been proven effective for noise cancellation or spectral filtering in a multitude of ways \cite{Dhar2021,Sekhar2022,Kato2022,Brochard2018,Fu2022,Pray2023}. Waves constructed from frequency combs contain all of their signal power in a few frequency peaks. So long as the wave interacts with a linear and stationary medium, no additional frequencies can be added to the true signal through interference. It is then possible to filter all but the chosen frequencies to reconstruct the signal. Noise will therefore adversely affect the signal $only$ where the frequency content of the noise is identical to these frequencies. This can reduce the impact of noise by orders of magnitude.

We therefore employ a method for constructing superoscillating functions using frequency combs to dramatically lessen the impact of noise. We follow the method developed in Refs.~\cite{karmakar2023beyond,aharonov2021new} to generate our superoscillating functions. This method has been shown capable of replicating any arbitrary function of finite length to arbitrary accuracy, even outside the signal's bandlimits. Briefly, we can mimic a function $f(t)$ over a finite interval $t \in \tau$ using $\psi(t)$ given by a finite sum of periodic functions 
\begin{equation}
    \psi(t) = \sum_{k=0}^{K-1} A_k e^{i \omega_{k} t},
    \label{psi}
\end{equation}
where the amplitudes $A_k$ are complex. For the realizations herein, we choose $N$ equally-spaced frequencies
\begin{equation}
    \omega_k = \frac{k}{K-1} \Omega + \omega_{min},
    \label{omega}
\end{equation} 
where $\omega_{min}$ is the minimum frequency and $\Omega$ is the frequency bandwidth. We can express  (\ref{psi}) in matrix notation 
\begin{equation}
    F=MA,
    \label{F}
\end{equation}
where $M$ is the matrix expression of $e^{i\omega_k t}$ for all sampled $k$ and $t$, $A$ is the vector of amplitude coefficients $A_k$, and $F$ is the function $f$ discretely sampled over $\tau$.  We can then take the Moore-Penrose pseudo-inverse matrix and solve for the amplitudes $A$.

This leads to a $\psi(t \in \tau)$ that is well-fit to $f(t \in \tau)$, even if $f$ utilizes a much larger bandwidth than the set of frequencies used for $\psi$. In trade, the lobes outside this "superregion" $\psi(t \notin \tau$) experience explosive growth in amplitude. This tradeoff demands either massive power outputs or aggressive noise reduction techniques to realize the superregion at any useful resolution.

Fortunately, the concentration of signal power allows for precise spectral filtering in post-processing. Hence, we envision the superoscillatory signal probing a stationary environment and then extracting the superoscillatory signal from the noisy returned signal. Using the method above, several pure frequencies of well-defined phases are summed to mimic any desired function in the superoscillating interval. The signal is generated and sent to probe a noisy medium. It is then measured. In postprocessing, the signal is transformed to the spectral domain, and all frequencies besides those of the original signal are set to zero. We then take the inverse Fourier transform to reconstruct the signal. 

In previous works we have proposed a novel method for range-resolving two reflectors given some prior assumptions \cite{Howell2023} and explored the mathematical limits of range resolution using several methods \cite{jordan2023fundamental}. We have further explored the ability to range-resolve two scatterers as a function of the structure of the outbound signal and proposed the ideal wave candidate to resolve two reflectors \cite{Jordan2024optimal}. In this paper, we apply superoscillation theory to the problem of range resolution while comparing our results to the range resolution limit formula (\ref{range_resolved}), which defines a figure of merit for unambiguous range resolution while making no prior assumptions about individual width, reflective amplitude, or exact number of scattering objects.

In the following sections, we will first provide a mathematical foundation for the benefit of frequency combs when spectrally filtering for noise reduction. Then, we show through in-lab experiments that even a wave whose superregion is buried in noise can be accurately reconstructed using precise spectral filtering, leading to sub-wavelength range resolution even in noisy environments. Lastly, we mention some important related findings.

\section{Frequency Combs for Noise Reduction}

Here we show that use of finite, discrete frequency series (frequency combs) leads to better signal-to-noise ratios for the frequencies of interest when dealing with white noise in the spectral domain. 

Any function $\psi(t)$ in the time domain can be converted to the frequency domain as $\Psi(\omega)$ by use of Fourier transform. For practical data processing, the orthonormal Fast Fourier Transform (FFT) of $\psi(t)$ is computed discretely and over a finite time interval as
\begin{equation}
    \Psi_{\omega} = \frac{1}{\sqrt{N}}\sum_{t=0}^{N-1} \psi_t e^{-\frac{2\pi t i}{N} \omega},
    \label{dft}
\end{equation}
where $N$ represents the total number of discrete time samples in the signal $\psi(t)$. The normalization factor of $\frac{1}{\sqrt{N}}$ assures that by Parseval's theorem, the total energy in the frequency domain is equal to the total energy in the time domain,
\begin{equation}
    \Big(E_{\Psi} = \sum_{k=0}^{N-1} \|\Psi_{\omega}\|^2\Big) = \Big(E_{\psi} = \sum_{t=0}^{N-1} \|\psi_t\|^2\Big).
    \label{parseval}
\end{equation}
One can then simplify (\ref{parseval}) to relate the average energy contribution from each frequency to the total energy and the number of sampled frequencies: 
\begin{equation}
    \langle E_{\Psi_{\omega}} \rangle = \frac{E_{\psi}}{N}. 
    \label{pwr}
\end{equation}
For all power signals (signals whose power in the time domain is finite but whose energy is infinite), the total energy of the sampled signal increases linearly with the number of samples ($E_{\psi} \propto N$). We can see, then, that for power signals with an infinite frequency spectrum, $\langle E_{\Psi_{\omega}} \rangle \propto N/N$ remains constant regardless of the sample size $N$. White noise is one such function.

\begin{figure*}[ht]
    \includegraphics[width=.95\textwidth]{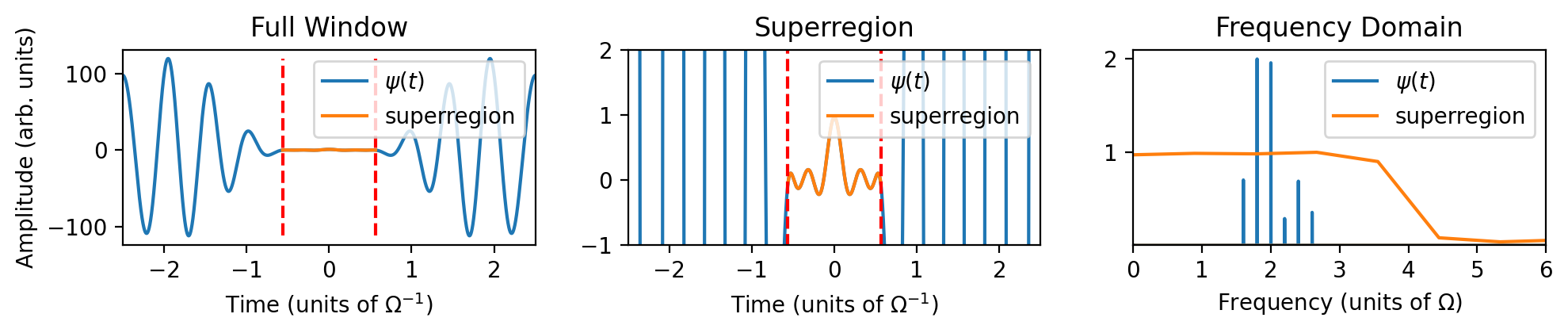}
    \caption{The constructed superfunction. (Left) and (middle) each show one complete cycle of the function in the time domain, zoomed differently in amplitude. (Right) shows the function in the frequency domain, both over the entire wavelength and just over the superregion (each scaled arbitrarily in amplitude for visual clarity). The full function contains a bandwidth ($\Omega$) only one-quarter of the bandwidth of the sinc wave the superfunction was fit to.}
    \label{waveform}    
\end{figure*}

For power signals that are bandlimited in the frequency domain, since frequencies outside the bandlimits do not contribute to the overall energy, the left side of (\ref{parseval}) becomes a sum of $k$ from 0 to $\kappa$ instead, where $\kappa$ is the number of sampled frequencies that fall within the bandlimits $\Omega$ of $\psi$. Then, $\langle E_{\Psi_{\omega}} \rangle \propto N/\kappa$. Should a bandlimited power signal's sampling rate increase in the time domain, $N$ increases but $\kappa$ does not. As white noise's spectral energy density is independent of $N$, increasing the sampling rate of a bandlimited power signal therefore increases $\Psi_{\omega \in \Omega}$'s signal-to-noise ratio (SNR) linearly with $N$. 

However, should the number of sampled wavelengths of $\psi$ increase but the sampling rate remain unaffected, $\Omega$ also increases. This leads to a $\kappa$ that scales linearly with $N$, resulting in zero net gain in the SNR of $\Psi_{\omega \in \Omega}$. A periodic frequency chirp would follow this pattern.

Frequency combs, whose frequency space is constructed of $K$ discrete frequencies (Eq. \ref{psi}), contain all of their energy in said $K$ frequencies entirely independent of $N$. Their spectral energy density becomes $\langle E_{\Psi_{\omega}} \rangle \propto N/K$. Since $K$ is a constant, $\langle E_{\Psi_{\omega}} \rangle$ will always grow linearly with $N$. This is independent of whether $N$ grows by increasing the number of time steps per cycle or by increasing the number of cycles.

This makes frequency combs unique among the family of power signals. Whereas any bandlimited power signal can improve its spectral SNR by increasing the sampling $rate$, only frequency combs can improve their SNR by increasing the size of the sampling $window$. Since it is much easier to collect multiple wave cycles than to increase the sampling rate, frequency combs can be leveraged to great effect during spectral filtering. 

\section{Analog Range Resolution Experiment}

\begin{figure}[ht]
    \includegraphics[width=.45\textwidth]{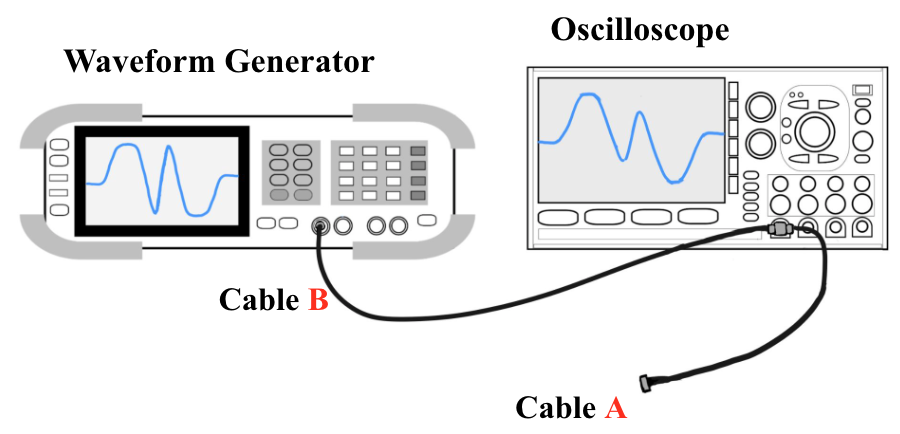}
    \caption{Experimental setup. A waveform generator sends a constructed pulse into a double-cable setup, such that the oscilloscope's collected signal is the sum of two waves that have traveled different total distances.}
    \label{exp_setup}    
\end{figure}

Here we show that it is possible to retrieve the superoscillating portion of a signal buried in noise in a real-world analog experiment using simple tools. A function generator is connected to an oscilloscope via BNC cables (Fig. \ref{exp_setup}). An extra cable length is connected to the oscilloscope end via T-junction and left open-ended. This creates two paths for the signal to travel down: one from the function generator to the scope, and one that travels down the extra cable length and back before terminating at the scope. This simulates a return signal scattered off of two point-like objects at different travel distances. 

We begin with a sinc function on the interval $[-4.5\pi, 4.5\pi]$. Sinc waves have the unique property of being a step function in the frequency domain, and are therefore convenient for this proof-of-concept.

We then construct a superfunction (Fig. \ref{waveform}) by defining $\omega_{k}$ in Eq. \ref{omega} such that $K=6,$ $\omega_{min}=.4\sigma,$ and $\Omega=.25\sigma,$ where $\sigma$ is the bandwidth of the sinc to fit to. Through testing, $K=6$ and $\omega_{min}=.4\sigma$ were found to be the best possible combination for all $\Omega=.25\sigma$ to minimize the lobe amplitudes while still keeping a high degree of accuracy for the fitted region (see supplemental material \cite{supp}). This results in a function that fits well to the sinc over the target region while containing a bandwidth only one-quarter of the sinc's. In trade, we see lobe amplitudes just over 100 times that of the maximum amplitude in the superregion.

The constructed superfunction was passed through the system in Fig. \ref{exp_setup} at a repetition rate of 400 kHz and a signal bandwidth of 2 MHz. Signal speed through the BNC cables was measured at $.65c$, where $c$ is the speed of light in free space. Every three meters of Cable A's length in Fig. \ref{exp_setup} therefore represents ~6.25\% of the function's inverse bandwidth in signal delay. The oscilloscope collected samples at a rate of 4 gigasamples/second, resulting in 10,000 sampled time steps per one complete wave cycle. 

The outbound signal's amplitude was set to 10 mVpp such that the natural noise in the experimental setup was enough to bury the superregion. The noise generated in the system was mostly white, but contained noticeable frequency-based attenuation that mildly affected our target frequencies. The overall signal's average power was calculated in postprocessing as 35 dB above the noise floor; the target superregion's average power was 17 dB below the noise floor. 

Cable A's length was then varied in three-meter (.0625 $\Omega^{-1}$) increments from 0 to 9 meters (0 to .1875 $\Omega^{-1}$) and 1,000 cycles ($10^{7}$ time steps) of each scattered return signal were saved. These were divided into ten-cycle increments for post-processing. Each ten-cycle increment was then converted to the frequency domain via FFT and all frequencies but the largest twelve (six positive and six negative) were set to 0. The signals were then returned to the time domain via IFFT and the quality of their reconstructions was evaluated.

\section{Results}

To quantitatively evaluate the reconstruction quality of the wave, we use the metric $MSE/E_{\psi}$, where $MSE$ is the evaluated mean squared error of the signal's reconstruction across multiple attempts. This normalizes the MSE relative to the total energy of the evaluated wave. 

\begin{figure}[h]
\includegraphics[width=.47\textwidth]{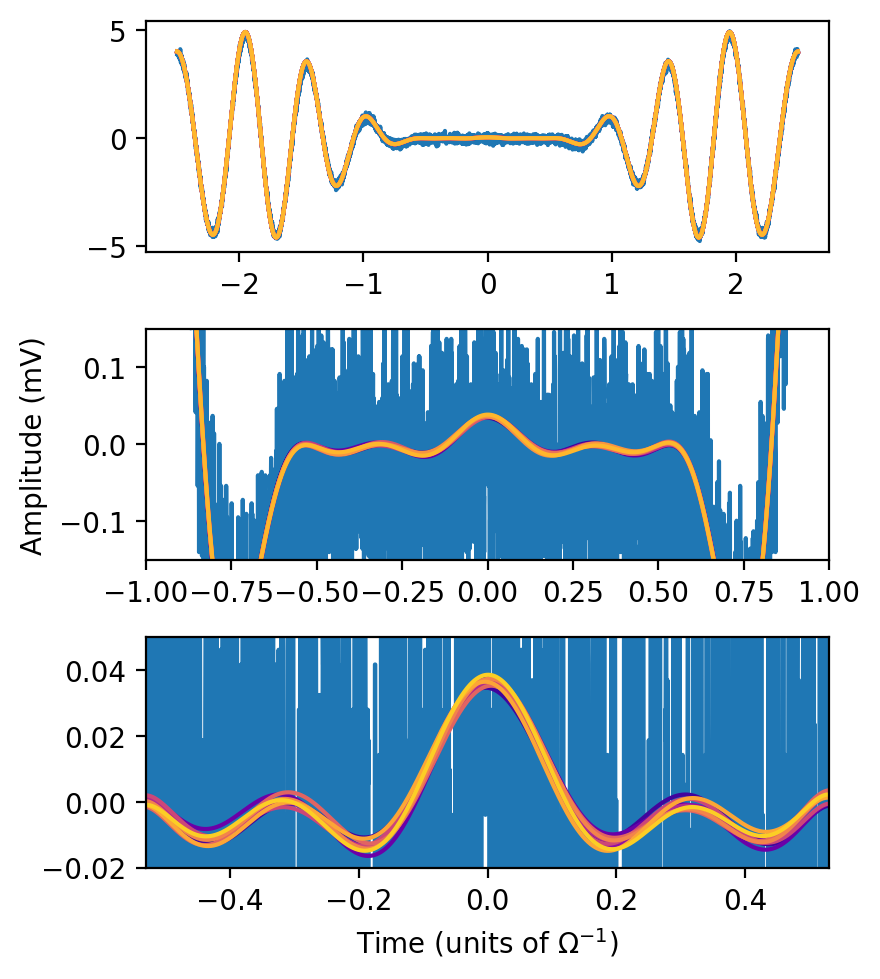}
    \caption{The full return signal (top). Despite being 17 dB below the noise floor (mid), the target superregion can still be reconstructed to high accuracy after spectrally filtering the FFT of ten cycles (bottom).} 
    \label{fig3}
\end{figure}

We find that the absolute MSE of the reconstructed wave is consistent throughout the wave's superregion and lobes. However, due to the very small signal amplitudes in the superregion, its normalized MSE is significantly larger than the normalized MSE of the total signal. Since it is the superregion we aim to reconstruct through spectral filtering, it is the superregion's normalized MSE values in particular that interest us. 

Using ten cycles of the return wave for each FFT, simple spectral noise elimination alone successfully reconstructed the target region to within a calculated mean squared error of 1.19\% of the superregion's total energy (Fig. \ref{fig3}). Using 99 cycles, this error drops to 0.16\%. 

\begin{figure}[h]
\includegraphics[width=.45\textwidth]{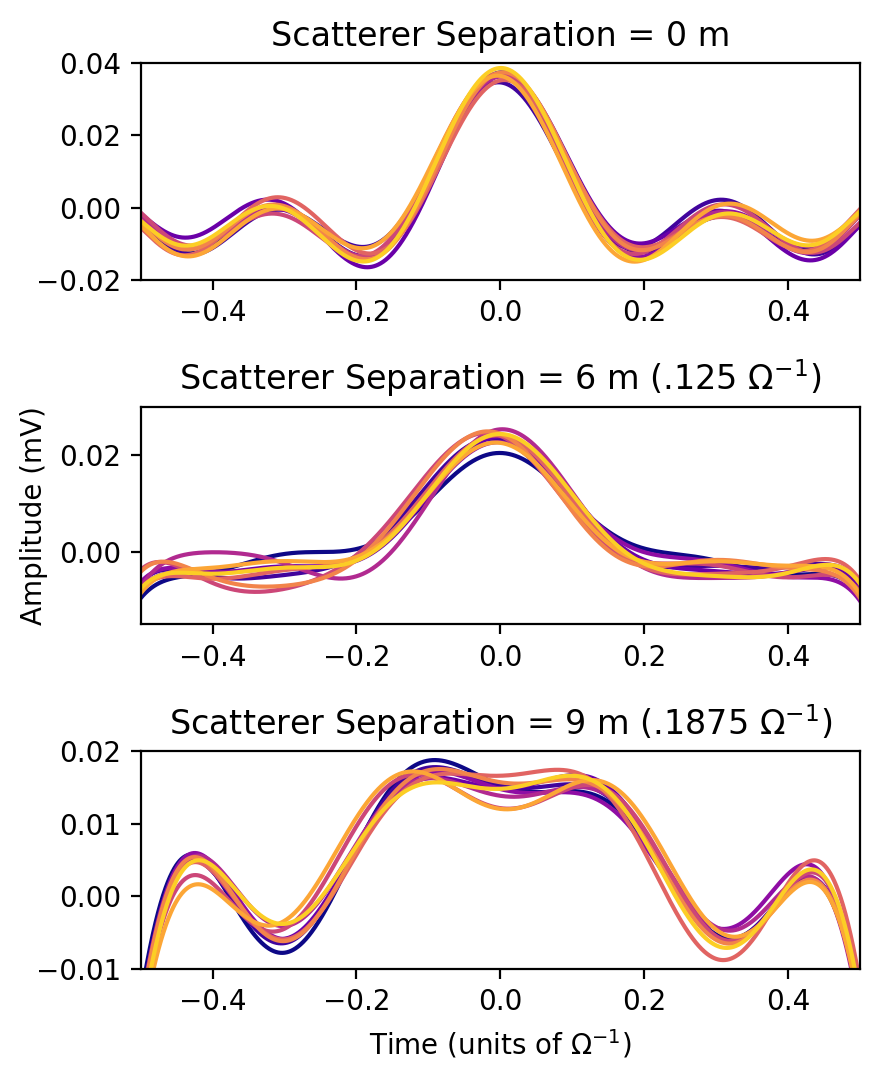}
    \caption{The superoscillating region, formerly buried in noise, is reconstructed to show the dramatic impact of scatterers less than a fifth of the inverse bandwidth apart. Ten reconstructions are shown, each by spectrally filtering ten cycles of the return wave, and each with independent and identically distributed noise realizations.} 
    \label{fig4}
\end{figure}

Further, since the superregion contains four times the frequency bandwidth of the total wave, the effects of changes in distance on the signal are amplified in this region. We therefore see significant and predictable distortion in the return signal's superregion from scatterer separations far below the inverse bandwidth. (Fig. \ref{fig4}). By just 0.1875 $\Omega^{-1}$, the signal has already separated enough to show two distinct scattering objects, meeting the range-resolved criterion well below the traditional range resolution limit of $\frac{1}{2 \Omega} $(Fig. \ref{fig4}, bottom).

\section{Discussion}

While neither frequency combs nor their benefits in noise reduction are new discoveries, their use in constructing superoscillating reproductions of any arbitrary analytic function is still relatively new. Here we have shown that their use also greatly aids in the highly publicized problem of noise when dealing with superoscillating functions. We have given a mathematical basis for their merit in spectral noise reduction. We have also shown experimentally that a superoscillating signal, despite being buried in noise, can nevertheless without ambiguity meet the Rayleigh criterion at distances that are impossible with ordinary signals given the frequencies used. 

It should be noted that despite the frequency-based attenuation found in the experimental setup, no effort was made to account for it here; additional spectral filtering techniques should help the signal reconstruction even further. Our only intent here is to show the efficacy of simple noise elimination with finite frequency series. We believe this method can easily be combined with other filtering methods to achieve greater results.

This method assumes that the return signal's sampling rate in frequency space is well-matched to the expected frequencies in the return signal (see supplemental materials \cite{supp}). Unaccounted for Doppler shifts, for example, severely impair the ability to accurately reconstruct the wave, more so than if the wave had been continuously non-zero within the bandlimits (non-combs). Existing probability-based techniques for dealing with Doppler shift should help ameliorate this.

Finally, it is worth noting that for simplicity, the bandwidth of our wave was considered as the extent of the signal's absolute bandlimits. This is a generous estimate, though; by either measuring the full width at half maximum (FWHM) or by calculating the standard deviation of the frequencies, the reported bandwidth of this signal would be much smaller. This means the criterion for range-resolved was likely met even farther below the inverse bandwidth than reported herein. 

\section{Acknowledgments}

JCH and ANJ acknowledge support from Chapman University and the Bill Hannon Foundation. AK acknowledges support through a Discovery Grant of the National Science and Engineering Council of Canada (NSERC), a Discovery Project grant of the Australian Research Council (ARC) and a Google Faculty Research Award. BS is supported by the Perimeter Institute, which is supported in part by the Government of Canada through the Department of Innovation, Science and Economic Development Canada and by the Province of Ontario through the Ministry of Economic Development, Job Creation and Trade.

\bibliography{Main}

\begin{thebibliography}{45}
\expandafter\ifx\csname natexlab\endcsname\relax\def\natexlab#1{#1}\fi
\expandafter\ifx\csname bibnamefont\endcsname\relax
  \def\bibnamefont#1{#1}\fi
\expandafter\ifx\csname bibfnamefont\endcsname\relax
  \def\bibfnamefont#1{#1}\fi
\expandafter\ifx\csname citenamefont\endcsname\relax
  \def\citenamefont#1{#1}\fi
\expandafter\ifx\csname url\endcsname\relax
  \def\url#1{\texttt{#1}}\fi
\expandafter\ifx\csname urlprefix\endcsname\relax\def\urlprefix{URL }\fi
\providecommand{\bibinfo}[2]{#2}
\providecommand{\eprint}[2][]{\url{#2}}

\bibitem[{\citenamefont{Aharonov et~al.}()\citenamefont{Aharonov, Popescu, and Rohrlich}}]{Aharonov-preprint}
\bibinfo{author}{\bibfnamefont{Y.}~\bibnamefont{Aharonov}}, \bibinfo{author}{\bibfnamefont{S.}~\bibnamefont{Popescu}}, \bibnamefont{and} \bibinfo{author}{\bibfnamefont{D.}~\bibnamefont{Rohrlich}}, \bibinfo{journal}{Tel-Aviv University Preprint TAUP} \textbf{\bibinfo{volume}{1847–90}} (????).

\bibitem[{\citenamefont{Popescu}(1991)}]{popescu1991multi}
\bibinfo{author}{\bibfnamefont{S.}~\bibnamefont{Popescu}}, Ph.D. thesis, \bibinfo{school}{PhD Thesis} (\bibinfo{year}{1991}).

\bibitem[{\citenamefont{Berry and Popescu}(2006)}]{berry2006evolution}
\bibinfo{author}{\bibfnamefont{M.}~\bibnamefont{Berry}} \bibnamefont{and} \bibinfo{author}{\bibfnamefont{S.}~\bibnamefont{Popescu}}, \bibinfo{journal}{Journal of Physics A: Mathematical and General} \textbf{\bibinfo{volume}{39}}, \bibinfo{pages}{6965} (\bibinfo{year}{2006}).

\bibitem[{\citenamefont{Kempf}(2018)}]{kempf2018four}
\bibinfo{author}{\bibfnamefont{A.}~\bibnamefont{Kempf}}, \bibinfo{journal}{Quantum Studies: Mathematics and Foundations} \textbf{\bibinfo{volume}{5}}, \bibinfo{pages}{477} (\bibinfo{year}{2018}).

\bibitem[{\citenamefont{Berry et~al.}(2019)\citenamefont{Berry, Zheludev, Aharonov, Colombo, Sabadini, Struppa, Tollaksen, Rogers, Qin, Hong et~al.}}]{berry2019roadmap}
\bibinfo{author}{\bibfnamefont{M.}~\bibnamefont{Berry}}, \bibinfo{author}{\bibfnamefont{N.}~\bibnamefont{Zheludev}}, \bibinfo{author}{\bibfnamefont{Y.}~\bibnamefont{Aharonov}}, \bibinfo{author}{\bibfnamefont{F.}~\bibnamefont{Colombo}}, \bibinfo{author}{\bibfnamefont{I.}~\bibnamefont{Sabadini}}, \bibinfo{author}{\bibfnamefont{D.~C.} \bibnamefont{Struppa}}, \bibinfo{author}{\bibfnamefont{J.}~\bibnamefont{Tollaksen}}, \bibinfo{author}{\bibfnamefont{E.~T.} \bibnamefont{Rogers}}, \bibinfo{author}{\bibfnamefont{F.}~\bibnamefont{Qin}}, \bibinfo{author}{\bibfnamefont{M.}~\bibnamefont{Hong}}, \bibnamefont{et~al.}, \bibinfo{journal}{Journal of Optics} \textbf{\bibinfo{volume}{21}}, \bibinfo{pages}{053002} (\bibinfo{year}{2019}).

\bibitem[{\citenamefont{Aharonov et~al.}(2011)\citenamefont{Aharonov, Colombo, Sabadini, Struppa, and Tollaksen}}]{Aharonov2011Aug}
\bibinfo{author}{\bibfnamefont{Y.}~\bibnamefont{Aharonov}}, \bibinfo{author}{\bibfnamefont{F.}~\bibnamefont{Colombo}}, \bibinfo{author}{\bibfnamefont{I.}~\bibnamefont{Sabadini}}, \bibinfo{author}{\bibfnamefont{D.~C.} \bibnamefont{Struppa}}, \bibnamefont{and} \bibinfo{author}{\bibfnamefont{J.}~\bibnamefont{Tollaksen}}, \bibinfo{journal}{J. Phys. A: Math. Theor.} \textbf{\bibinfo{volume}{44}}, \bibinfo{pages}{365304} (\bibinfo{year}{2011}), ISSN \bibinfo{issn}{1751-8113}.

\bibitem[{\citenamefont{Dressel et~al.}(2014)\citenamefont{Dressel, Malik, Miatto, Jordan, and Boyd}}]{dressel2014colloquium}
\bibinfo{author}{\bibfnamefont{J.}~\bibnamefont{Dressel}}, \bibinfo{author}{\bibfnamefont{M.}~\bibnamefont{Malik}}, \bibinfo{author}{\bibfnamefont{F.~M.} \bibnamefont{Miatto}}, \bibinfo{author}{\bibfnamefont{A.~N.} \bibnamefont{Jordan}}, \bibnamefont{and} \bibinfo{author}{\bibfnamefont{R.~W.} \bibnamefont{Boyd}}, \bibinfo{journal}{Reviews of Modern Physics} \textbf{\bibinfo{volume}{86}}, \bibinfo{pages}{307} (\bibinfo{year}{2014}).

\bibitem[{\citenamefont{Jordan et~al.}(2022)\citenamefont{Jordan, Aharonov, Struppa, Colombo, Sabadini, Shushi, Tollaksen, Howell, and Vamivakas}}]{jordan2022super}
\bibinfo{author}{\bibfnamefont{A.~N.} \bibnamefont{Jordan}}, \bibinfo{author}{\bibfnamefont{Y.}~\bibnamefont{Aharonov}}, \bibinfo{author}{\bibfnamefont{D.~C.} \bibnamefont{Struppa}}, \bibinfo{author}{\bibfnamefont{F.}~\bibnamefont{Colombo}}, \bibinfo{author}{\bibfnamefont{I.}~\bibnamefont{Sabadini}}, \bibinfo{author}{\bibfnamefont{T.}~\bibnamefont{Shushi}}, \bibinfo{author}{\bibfnamefont{J.}~\bibnamefont{Tollaksen}}, \bibinfo{author}{\bibfnamefont{J.~C.} \bibnamefont{Howell}}, \bibnamefont{and} \bibinfo{author}{\bibfnamefont{A.~N.} \bibnamefont{Vamivakas}}, \bibinfo{journal}{arXiv preprint arXiv:2209.05650}  (\bibinfo{year}{2022}).

\bibitem[{\citenamefont{Jordan and Siddiqi}(2024)}]{jordan2024quantum}
\bibinfo{author}{\bibfnamefont{A.~N.} \bibnamefont{Jordan}} \bibnamefont{and} \bibinfo{author}{\bibfnamefont{I.~A.} \bibnamefont{Siddiqi}}, \emph{\bibinfo{title}{Quantum Measurement: Theory and Practice}} (\bibinfo{publisher}{Cambridge University Press}, \bibinfo{year}{2024}).

\bibitem[{\citenamefont{Huang and Zheludev}(2009)}]{huang2009super}
\bibinfo{author}{\bibfnamefont{F.~M.} \bibnamefont{Huang}} \bibnamefont{and} \bibinfo{author}{\bibfnamefont{N.~I.} \bibnamefont{Zheludev}}, \bibinfo{journal}{Nano letters} \textbf{\bibinfo{volume}{9}}, \bibinfo{pages}{1249} (\bibinfo{year}{2009}).

\bibitem[{\citenamefont{Rogers et~al.}(2012)\citenamefont{Rogers, Lindberg, Roy, Savo, Chad, Dennis, and Zheludev}}]{rogers2012super}
\bibinfo{author}{\bibfnamefont{E.~T.} \bibnamefont{Rogers}}, \bibinfo{author}{\bibfnamefont{J.}~\bibnamefont{Lindberg}}, \bibinfo{author}{\bibfnamefont{T.}~\bibnamefont{Roy}}, \bibinfo{author}{\bibfnamefont{S.}~\bibnamefont{Savo}}, \bibinfo{author}{\bibfnamefont{J.~E.} \bibnamefont{Chad}}, \bibinfo{author}{\bibfnamefont{M.~R.} \bibnamefont{Dennis}}, \bibnamefont{and} \bibinfo{author}{\bibfnamefont{N.~I.} \bibnamefont{Zheludev}}, \bibinfo{journal}{Nature materials} \textbf{\bibinfo{volume}{11}}, \bibinfo{pages}{432} (\bibinfo{year}{2012}).

\bibitem[{\citenamefont{Kozawa et~al.}(2018)\citenamefont{Kozawa, Matsunaga, and Sato}}]{Kozawa2018Feb}
\bibinfo{author}{\bibfnamefont{Y.}~\bibnamefont{Kozawa}}, \bibinfo{author}{\bibfnamefont{D.}~\bibnamefont{Matsunaga}}, \bibnamefont{and} \bibinfo{author}{\bibfnamefont{S.}~\bibnamefont{Sato}}, \bibinfo{journal}{Optica} \textbf{\bibinfo{volume}{5}}, \bibinfo{pages}{86} (\bibinfo{year}{2018}), ISSN \bibinfo{issn}{2334-2536}.

\bibitem[{\citenamefont{Gbur}(2019)}]{gbur2019using}
\bibinfo{author}{\bibfnamefont{G.}~\bibnamefont{Gbur}}, \bibinfo{journal}{Nanophotonics} \textbf{\bibinfo{volume}{8}}, \bibinfo{pages}{205} (\bibinfo{year}{2019}).

\bibitem[{\citenamefont{Wong and Eleftheriades}(2013)}]{Wong2013}
\bibinfo{author}{\bibfnamefont{A.~M.~H.} \bibnamefont{Wong}} \bibnamefont{and} \bibinfo{author}{\bibfnamefont{G.~V.} \bibnamefont{Eleftheriades}}, \bibinfo{journal}{Scientific Reports} \textbf{\bibinfo{volume}{3}}, \bibinfo{pages}{1715} (\bibinfo{year}{2013}), ISSN \bibinfo{issn}{2045-2322}.

\bibitem[{\citenamefont{Hao et~al.}(2017)\citenamefont{Hao, Nie, Ye, Li, Luo, Feng, Yu, Wen, Zhang, Yu et~al.}}]{hao2017three}
\bibinfo{author}{\bibfnamefont{C.}~\bibnamefont{Hao}}, \bibinfo{author}{\bibfnamefont{Z.}~\bibnamefont{Nie}}, \bibinfo{author}{\bibfnamefont{H.}~\bibnamefont{Ye}}, \bibinfo{author}{\bibfnamefont{H.}~\bibnamefont{Li}}, \bibinfo{author}{\bibfnamefont{Y.}~\bibnamefont{Luo}}, \bibinfo{author}{\bibfnamefont{R.}~\bibnamefont{Feng}}, \bibinfo{author}{\bibfnamefont{X.}~\bibnamefont{Yu}}, \bibinfo{author}{\bibfnamefont{F.}~\bibnamefont{Wen}}, \bibinfo{author}{\bibfnamefont{Y.}~\bibnamefont{Zhang}}, \bibinfo{author}{\bibfnamefont{C.}~\bibnamefont{Yu}}, \bibnamefont{et~al.}, \bibinfo{journal}{Science advances} \textbf{\bibinfo{volume}{3}}, \bibinfo{pages}{e1701398} (\bibinfo{year}{2017}).

\bibitem[{\citenamefont{Eliezer et~al.}(2017)\citenamefont{Eliezer, Hareli, Lobachinsky, Froim, and Bahabad}}]{eliezer2017breaking}
\bibinfo{author}{\bibfnamefont{Y.}~\bibnamefont{Eliezer}}, \bibinfo{author}{\bibfnamefont{L.}~\bibnamefont{Hareli}}, \bibinfo{author}{\bibfnamefont{L.}~\bibnamefont{Lobachinsky}}, \bibinfo{author}{\bibfnamefont{S.}~\bibnamefont{Froim}}, \bibnamefont{and} \bibinfo{author}{\bibfnamefont{A.}~\bibnamefont{Bahabad}}, \bibinfo{journal}{Physical review letters} \textbf{\bibinfo{volume}{119}}, \bibinfo{pages}{043903} (\bibinfo{year}{2017}).

\bibitem[{\citenamefont{Makris and Psaltis}(2011)}]{makris2011superoscillatory}
\bibinfo{author}{\bibfnamefont{K.~G.} \bibnamefont{Makris}} \bibnamefont{and} \bibinfo{author}{\bibfnamefont{D.}~\bibnamefont{Psaltis}}, \bibinfo{journal}{Optics letters} \textbf{\bibinfo{volume}{36}}, \bibinfo{pages}{4335} (\bibinfo{year}{2011}).

\bibitem[{\citenamefont{Greenfield et~al.}(2013)\citenamefont{Greenfield, Schley, Hurwitz, Nemirovsky, Makris, and Segev}}]{greenfield2013experimental}
\bibinfo{author}{\bibfnamefont{E.}~\bibnamefont{Greenfield}}, \bibinfo{author}{\bibfnamefont{R.}~\bibnamefont{Schley}}, \bibinfo{author}{\bibfnamefont{I.}~\bibnamefont{Hurwitz}}, \bibinfo{author}{\bibfnamefont{J.}~\bibnamefont{Nemirovsky}}, \bibinfo{author}{\bibfnamefont{K.~G.} \bibnamefont{Makris}}, \bibnamefont{and} \bibinfo{author}{\bibfnamefont{M.}~\bibnamefont{Segev}}, \bibinfo{journal}{Optics Express} \textbf{\bibinfo{volume}{21}}, \bibinfo{pages}{13425} (\bibinfo{year}{2013}).

\bibitem[{\citenamefont{Yuan et~al.}(2017)\citenamefont{Yuan, Rogers, and Zheludev}}]{yuan2017achromatic}
\bibinfo{author}{\bibfnamefont{G.~H.} \bibnamefont{Yuan}}, \bibinfo{author}{\bibfnamefont{E.~T.} \bibnamefont{Rogers}}, \bibnamefont{and} \bibinfo{author}{\bibfnamefont{N.~I.} \bibnamefont{Zheludev}}, \bibinfo{journal}{Light: Science \& Applications} \textbf{\bibinfo{volume}{6}}, \bibinfo{pages}{e17036} (\bibinfo{year}{2017}).

\bibitem[{\citenamefont{Wong and Eleftheriades}(2011{\natexlab{a}})}]{wong2011b}
\bibinfo{author}{\bibfnamefont{A.~M.~H.} \bibnamefont{Wong}} \bibnamefont{and} \bibinfo{author}{\bibfnamefont{G.~V.} \bibnamefont{Eleftheriades}}, \bibinfo{journal}{IEEE Transactions on Microwave Theory and Techniques} \textbf{\bibinfo{volume}{59}}, \bibinfo{pages}{2173} (\bibinfo{year}{2011}{\natexlab{a}}).

\bibitem[{\citenamefont{Wong and Eleftheriades}(2017)}]{Wong2017}
\bibinfo{author}{\bibfnamefont{A.~M.~H.} \bibnamefont{Wong}} \bibnamefont{and} \bibinfo{author}{\bibfnamefont{G.~V.} \bibnamefont{Eleftheriades}}, \bibinfo{journal}{Physical Review B} \textbf{\bibinfo{volume}{95}}, \bibinfo{pages}{075148} (\bibinfo{year}{2017}), ISSN \bibinfo{issn}{2469-9950, 2469-9969}.

\bibitem[{\citenamefont{Yuan et~al.}(2014)\citenamefont{Yuan, Rogers, Roy, Shen, and Zheludev}}]{Yuan2014}
\bibinfo{author}{\bibfnamefont{G.}~\bibnamefont{Yuan}}, \bibinfo{author}{\bibfnamefont{E.~T.~F.} \bibnamefont{Rogers}}, \bibinfo{author}{\bibfnamefont{T.}~\bibnamefont{Roy}}, \bibinfo{author}{\bibfnamefont{Z.}~\bibnamefont{Shen}}, \bibnamefont{and} \bibinfo{author}{\bibfnamefont{N.~I.} \bibnamefont{Zheludev}}, \bibinfo{journal}{Optics Express} \textbf{\bibinfo{volume}{22}}, \bibinfo{pages}{6428} (\bibinfo{year}{2014}), ISSN \bibinfo{issn}{1094-4087}.

\bibitem[{\citenamefont{Xiao et~al.}(2024)\citenamefont{Xiao, Xiong, Wang, Zhou, Dai, Cai, and Yan}}]{Xiao2024}
\bibinfo{author}{\bibfnamefont{R.}~\bibnamefont{Xiao}}, \bibinfo{author}{\bibfnamefont{J.}~\bibnamefont{Xiong}}, \bibinfo{author}{\bibfnamefont{Z.}~\bibnamefont{Wang}}, \bibinfo{author}{\bibfnamefont{J.}~\bibnamefont{Zhou}}, \bibinfo{author}{\bibfnamefont{G.}~\bibnamefont{Dai}}, \bibinfo{author}{\bibfnamefont{M.}~\bibnamefont{Cai}}, \bibnamefont{and} \bibinfo{author}{\bibfnamefont{W.}~\bibnamefont{Yan}}, \bibinfo{journal}{Journal of Optics} \textbf{\bibinfo{volume}{26}} (\bibinfo{year}{2024}), ISSN \bibinfo{issn}{2040-8978, 2040-8986}.

\bibitem[{\citenamefont{Berry}(1994)}]{berry1994faster}
\bibinfo{author}{\bibfnamefont{M.}~\bibnamefont{Berry}}, \emph{\bibinfo{title}{Faster than {F}ourier {Q}uantum {C}oherence and {R}eality; in celebration of the 60th {B}irthday of {Y}akir {A}haronov ed {JS} {A}nandan and {JL} {S}afko}} (\bibinfo{year}{1994}).

\bibitem[{\citenamefont{Wong and Eleftheriades}(2011{\natexlab{b}})}]{Wong2011}
\bibinfo{author}{\bibfnamefont{A.~M.~H.} \bibnamefont{Wong}} \bibnamefont{and} \bibinfo{author}{\bibfnamefont{G.~V.} \bibnamefont{Eleftheriades}}, \bibinfo{journal}{IEEE Transactions on Antennas and Propagation} \textbf{\bibinfo{volume}{59}}, \bibinfo{pages}{4766} (\bibinfo{year}{2011}{\natexlab{b}}).

\bibitem[{\citenamefont{Wong and Eleftheriades}(2012)}]{wong2012superoscillatory}
\bibinfo{author}{\bibfnamefont{A.~M.} \bibnamefont{Wong}} \bibnamefont{and} \bibinfo{author}{\bibfnamefont{G.~V.} \bibnamefont{Eleftheriades}}, \bibinfo{journal}{IEEE microwave and wireless components letters} \textbf{\bibinfo{volume}{22}}, \bibinfo{pages}{147} (\bibinfo{year}{2012}).

\bibitem[{\citenamefont{Wehner}(1995)}]{Wehner_1995}
\bibinfo{author}{\bibfnamefont{D.~R.} \bibnamefont{Wehner}}, \emph{\bibinfo{title}{High-resolution radar}} (\bibinfo{publisher}{Artech House}, \bibinfo{address}{Boston London}, \bibinfo{year}{1995}), \bibinfo{edition}{2nd} ed., ISBN \bibinfo{isbn}{978-0-89006-727-7}.

\bibitem[{\citenamefont{Nuss et~al.}(2020)\citenamefont{Nuss, Mayer, Marahrens, and Zwick}}]{Nuss2020}
\bibinfo{author}{\bibfnamefont{B.}~\bibnamefont{Nuss}}, \bibinfo{author}{\bibfnamefont{J.}~\bibnamefont{Mayer}}, \bibinfo{author}{\bibfnamefont{S.}~\bibnamefont{Marahrens}}, \bibnamefont{and} \bibinfo{author}{\bibfnamefont{T.}~\bibnamefont{Zwick}}, \bibinfo{journal}{IEEE Transactions on Microwave Theory and Techniques} \textbf{\bibinfo{volume}{68}}, \bibinfo{pages}{3861–3871} (\bibinfo{year}{2020}), ISSN \bibinfo{issn}{0018-9480, 1557-9670}.

\bibitem[{\citenamefont{Sengupta et~al.}(2024)\citenamefont{Sengupta, Grzeslo, Iwamatsu, Tebart, Haddad, and Stöhr}}]{Sengupta2024}
\bibinfo{author}{\bibfnamefont{N.}~\bibnamefont{Sengupta}}, \bibinfo{author}{\bibfnamefont{M.}~\bibnamefont{Grzeslo}}, \bibinfo{author}{\bibfnamefont{S.}~\bibnamefont{Iwamatsu}}, \bibinfo{author}{\bibfnamefont{J.}~\bibnamefont{Tebart}}, \bibinfo{author}{\bibfnamefont{T.}~\bibnamefont{Haddad}}, \bibnamefont{and} \bibinfo{author}{\bibfnamefont{A.}~\bibnamefont{Stöhr}}, in \emph{\bibinfo{booktitle}{Proceedings of the 4th URSI Atlantic RadioScience Conference – AT-RASC 2024}} (\bibinfo{publisher}{URSI – International Union of Radio Science}, \bibinfo{address}{Gran Canaria, Spain}, \bibinfo{year}{2024}), ISBN \bibinfo{isbn}{978-94-6396-810-2}.

\bibitem[{\citenamefont{Lang et~al.}(2020)\citenamefont{Lang, Onic, Schmid, Feger, and Huemer}}]{Lang2020}
\bibinfo{author}{\bibfnamefont{O.}~\bibnamefont{Lang}}, \bibinfo{author}{\bibfnamefont{A.}~\bibnamefont{Onic}}, \bibinfo{author}{\bibfnamefont{C.}~\bibnamefont{Schmid}}, \bibinfo{author}{\bibfnamefont{R.}~\bibnamefont{Feger}}, \bibnamefont{and} \bibinfo{author}{\bibfnamefont{M.}~\bibnamefont{Huemer}}, in \emph{\bibinfo{booktitle}{2020 54th Asilomar Conference on Signals, Systems, and Computers}} (\bibinfo{publisher}{IEEE}, \bibinfo{address}{Pacific Grove, CA, USA}, \bibinfo{year}{2020}), p. \bibinfo{pages}{1563–1567}, ISBN \bibinfo{isbn}{978-0-7381-3126-9}.

\bibitem[{\citenamefont{Vasconcelos et~al.}(2023)\citenamefont{Vasconcelos, Nallabolu, and Li}}]{Vasconcelos2023}
\bibinfo{author}{\bibfnamefont{M.}~\bibnamefont{Vasconcelos}}, \bibinfo{author}{\bibfnamefont{P.}~\bibnamefont{Nallabolu}}, \bibnamefont{and} \bibinfo{author}{\bibfnamefont{C.}~\bibnamefont{Li}}, in \emph{\bibinfo{booktitle}{2023 IEEE Topical Conference on Wireless Sensors and Sensor Networks}} (\bibinfo{publisher}{IEEE}, \bibinfo{address}{Las Vegas, NV, USA}, \bibinfo{year}{2023}), p. \bibinfo{pages}{53–56}, ISBN \bibinfo{isbn}{978-1-66549-321-5}.

\bibitem[{\citenamefont{Ferreira and Kempf}(2002)}]{EnergyExpense}
\bibinfo{author}{\bibfnamefont{P.~J. S.~G.} \bibnamefont{Ferreira}} \bibnamefont{and} \bibinfo{author}{\bibfnamefont{A.}~\bibnamefont{Kempf}}, in \emph{\bibinfo{booktitle}{2002 11th European Signal Processing Conference}} (\bibinfo{year}{2002}), pp. \bibinfo{pages}{1--4}.

\bibitem[{\citenamefont{Berry}(2016)}]{Berry_2017}
\bibinfo{author}{\bibfnamefont{M.~V.} \bibnamefont{Berry}}, \bibinfo{journal}{Journal of Physics A: Mathematical and Theoretical} \textbf{\bibinfo{volume}{50}}, \bibinfo{pages}{025003} (\bibinfo{year}{2016}).

\bibitem[{\citenamefont{Dhar et~al.}(2021)\citenamefont{Dhar, Deyasi, and Sarkar}}]{Dhar2021}
\bibinfo{author}{\bibfnamefont{R.}~\bibnamefont{Dhar}}, \bibinfo{author}{\bibfnamefont{A.}~\bibnamefont{Deyasi}}, \bibnamefont{and} \bibinfo{author}{\bibfnamefont{A.}~\bibnamefont{Sarkar}}, in \emph{\bibinfo{booktitle}{2021 Devices for Integrated Circuit (DevIC)}} (\bibinfo{year}{2021}), pp. \bibinfo{pages}{13--18}.

\bibitem[{\citenamefont{Sekhar et~al.}(2022)\citenamefont{Sekhar, Fredrick, Wu, Swartz, and Diddams}}]{Sekhar2022}
\bibinfo{author}{\bibfnamefont{P.}~\bibnamefont{Sekhar}}, \bibinfo{author}{\bibfnamefont{C.}~\bibnamefont{Fredrick}}, \bibinfo{author}{\bibfnamefont{T.-H.} \bibnamefont{Wu}}, \bibinfo{author}{\bibfnamefont{S.}~\bibnamefont{Swartz}}, \bibnamefont{and} \bibinfo{author}{\bibfnamefont{S.~A.} \bibnamefont{Diddams}}, in \emph{\bibinfo{booktitle}{2022 IEEE Photonics Conference (IPC)}} (\bibinfo{year}{2022}), pp. \bibinfo{pages}{1--2}.

\bibitem[{\citenamefont{Kato et~al.}(2022)\citenamefont{Kato, Morito, Nekoshima, and Minoshima}}]{Kato2022}
\bibinfo{author}{\bibfnamefont{T.}~\bibnamefont{Kato}}, \bibinfo{author}{\bibfnamefont{T.}~\bibnamefont{Morito}}, \bibinfo{author}{\bibfnamefont{Y.}~\bibnamefont{Nekoshima}}, \bibnamefont{and} \bibinfo{author}{\bibfnamefont{K.}~\bibnamefont{Minoshima}}, in \emph{\bibinfo{booktitle}{2022 Conference on Lasers and Electro-Optics Pacific Rim (CLEO-PR)}} (\bibinfo{year}{2022}), pp. \bibinfo{pages}{1--2}.

\bibitem[{\citenamefont{Brochard et~al.}(2018)\citenamefont{Brochard, Schilt, and Südmeyer}}]{Brochard2018}
\bibinfo{author}{\bibfnamefont{P.}~\bibnamefont{Brochard}}, \bibinfo{author}{\bibfnamefont{S.}~\bibnamefont{Schilt}}, \bibnamefont{and} \bibinfo{author}{\bibfnamefont{T.}~\bibnamefont{Südmeyer}}, in \emph{\bibinfo{booktitle}{2018 International Topical Meeting on Microwave Photonics (MWP)}} (\bibinfo{year}{2018}), pp. \bibinfo{pages}{1--4}.

\bibitem[{\citenamefont{Fu et~al.}(2022)\citenamefont{Fu, Lu, Yan, Deng, Zhu, and Wang}}]{Fu2022}
\bibinfo{author}{\bibfnamefont{F.}~\bibnamefont{Fu}}, \bibinfo{author}{\bibfnamefont{B.}~\bibnamefont{Lu}}, \bibinfo{author}{\bibfnamefont{X.}~\bibnamefont{Yan}}, \bibinfo{author}{\bibfnamefont{M.}~\bibnamefont{Deng}}, \bibinfo{author}{\bibfnamefont{L.}~\bibnamefont{Zhu}}, \bibnamefont{and} \bibinfo{author}{\bibfnamefont{A.}~\bibnamefont{Wang}}, in \emph{\bibinfo{booktitle}{2022 IEEE International Topical Meeting on Microwave Photonics (MWP)}} (\bibinfo{year}{2022}), pp. \bibinfo{pages}{1--4}.

\bibitem[{\citenamefont{Prayoonyong and Corcoran}(2023)}]{Pray2023}
\bibinfo{author}{\bibfnamefont{C.}~\bibnamefont{Prayoonyong}} \bibnamefont{and} \bibinfo{author}{\bibfnamefont{B.}~\bibnamefont{Corcoran}}, \bibinfo{journal}{Journal of Lightwave Technology} \textbf{\bibinfo{volume}{41}}, \bibinfo{pages}{3034} (\bibinfo{year}{2023}).

\bibitem[{\citenamefont{Karmakar and Jordan}(2023)}]{karmakar2023beyond}
\bibinfo{author}{\bibfnamefont{T.}~\bibnamefont{Karmakar}} \bibnamefont{and} \bibinfo{author}{\bibfnamefont{A.~N.} \bibnamefont{Jordan}}, \bibinfo{journal}{Journal of Physics A: Mathematical and Theoretical} \textbf{\bibinfo{volume}{56}}, \bibinfo{pages}{495204} (\bibinfo{year}{2023}).

\bibitem[{\citenamefont{Aharonov et~al.}(2021)\citenamefont{Aharonov, Colombo, Sabadini, Shushi, Struppa, and Tollaksen}}]{aharonov2021new}
\bibinfo{author}{\bibfnamefont{Y.}~\bibnamefont{Aharonov}}, \bibinfo{author}{\bibfnamefont{F.}~\bibnamefont{Colombo}}, \bibinfo{author}{\bibfnamefont{I.}~\bibnamefont{Sabadini}}, \bibinfo{author}{\bibfnamefont{T.}~\bibnamefont{Shushi}}, \bibinfo{author}{\bibfnamefont{D.~C.} \bibnamefont{Struppa}}, \bibnamefont{and} \bibinfo{author}{\bibfnamefont{J.}~\bibnamefont{Tollaksen}}, \bibinfo{journal}{Proceedings of the Royal Society A} \textbf{\bibinfo{volume}{477}}, \bibinfo{pages}{20210020} (\bibinfo{year}{2021}).

\bibitem[{\citenamefont{Howell et~al.}(2023)\citenamefont{Howell, Jordan, \ifmmode~\check{S}\else \v{S}\fi{}oda, and Kempf}}]{Howell2023}
\bibinfo{author}{\bibfnamefont{J.~C.} \bibnamefont{Howell}}, \bibinfo{author}{\bibfnamefont{A.~N.} \bibnamefont{Jordan}}, \bibinfo{author}{\bibfnamefont{B.}~\bibnamefont{\ifmmode~\check{S}\else \v{S}\fi{}oda}}, \bibnamefont{and} \bibinfo{author}{\bibfnamefont{A.}~\bibnamefont{Kempf}}, \bibinfo{journal}{Phys. Rev. Lett.} \textbf{\bibinfo{volume}{131}}, \bibinfo{pages}{053803} (\bibinfo{year}{2023}).

\bibitem[{\citenamefont{Jordan and Howell}(2023)}]{jordan2023fundamental}
\bibinfo{author}{\bibfnamefont{A.~N.} \bibnamefont{Jordan}} \bibnamefont{and} \bibinfo{author}{\bibfnamefont{J.~C.} \bibnamefont{Howell}}, \bibinfo{journal}{Physical Review Applied} \textbf{\bibinfo{volume}{20}}, \bibinfo{pages}{064046} (\bibinfo{year}{2023}).

\bibitem[{\citenamefont{Jordan et~al.}(2024)\citenamefont{Jordan, Howell, Kempf, Zhang, and White}}]{Jordan2024optimal}
\bibinfo{author}{\bibfnamefont{A.~N.} \bibnamefont{Jordan}}, \bibinfo{author}{\bibfnamefont{J.~C.} \bibnamefont{Howell}}, \bibinfo{author}{\bibfnamefont{A.}~\bibnamefont{Kempf}}, \bibinfo{author}{\bibfnamefont{S.}~\bibnamefont{Zhang}}, \bibnamefont{and} \bibinfo{author}{\bibfnamefont{D.}~\bibnamefont{White}}, \bibinfo{journal}{Physical Review Research} \textbf{\bibinfo{volume}{6}} (\bibinfo{year}{2024}), ISSN \bibinfo{issn}{2643-1564}.

\bibitem[{sup()}]{supp}
\emph{\bibinfo{title}{See supplemental material at [url] for details on superfunction parameter selection to minimize lobe amplitudes while maintaining high fitting accuracy.}}

\end{thebibliography}
\clearpage
\appendix
\onecolumngrid





\section{Supplemental Material}

The goal here is, given Eqs. (1) and (2), to find a set of frequencies that fulfills the seemingly contradictory tasks of both minimizing the error between $\psi(t)$ and $f(t)$ over some interval $\tau$ and minimizing the amplitude of $\psi$'s lobes outside $\tau$. To accomplish this, we set up a simple grid search for $K$, $\Omega$, and $\omega_{min}$, as follows:

\begin{algorithm}
\caption{GridSearch to Maximize Fit and Minimize Lobes}
\begin{algorithmic}[1]
\State \textbf{Input:} array $\tau$: evenly-spaced timesteps defining the finite interval over which to fit the function $\psi$
\State \textbf{Input:} array $f(\tau)$: the target function to fit to, sampled at $\tau$'s timesteps
\State \textbf{Output:} array $R$, where $R_{i}$ = [K$_i$, $\Omega_i$, $\omega_{min_i}$, $\psi_{max_i}$, MSE$_i$]
    \For{$\Omega \gets 0.1$ to $0.5$ step $0.05$}
        \For{$\omega_{min} \gets 0.0$ to $1-\Omega$ step $0.01$}
            \For{$K \gets 3$ to $20$}
                \For{$k \gets 0$ to $K-1$}
                    \State $\omega_{k}=\frac{k}{K-1}\Omega+\omega_{min}$
                \EndFor
                \State $\psi(t) = \sum_{k=0}^{K-1} A_k e^{i \omega_{k} t}$
                \State $A = M^{-1}F$
                \State $\psi_{max}$ = max($\psi(t)$) 
                \State MSE = mean($(\psi(\tau) - f(\tau))^2$)
                \State append [K, $\Omega$, $\omega_{min}$, $\psi_{max}$, MSE] to R
            \EndFor
        \EndFor
    \EndFor
    \State \textbf{return} $R$
\end{algorithmic}
\end{algorithm}

From here, we can view the impact of $K$, $\Omega$, and $\omega_{min}$ on $\psi_{max}$ and $MSE$ by filtering and plotting along any of the axes $K$, $\Omega$, and $\omega_{min}$. It is important to note that the relationships found between the above variables are specific to the process of fitting a frequency comb to a sinc wave, whose frequency content is equally distributed and continuous over a finite bandlimit. It serves as a line of later investigation to see how many of these fitting principles generalize to fitting to other waveform types as well.

\begin{figure}[H]
    \centering
    \includegraphics[width=.95\textwidth]{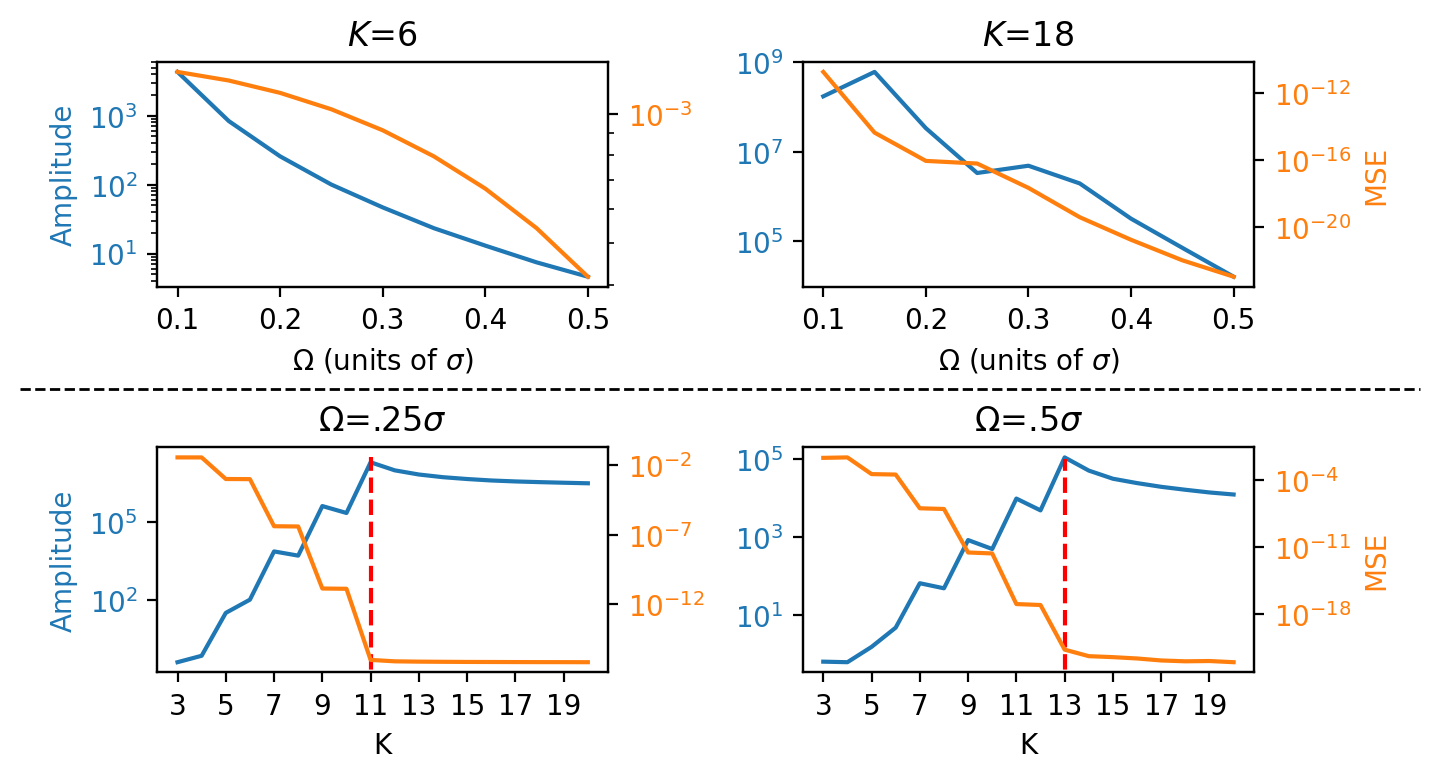}
    \caption{Amplitude and MSE as functions of $\Omega$ for fixed $K$ (top) and of $K$ for fixed $\Omega$ (bottom). The dashed line in (bottom) indicates $K^* = 11$ for $\Omega = .25\sigma$ and $K^* = 13$ for $\Omega = .5\sigma$, at which point both the amplitude and MSE plateau. These patterns hold true for any $K$ and $\Omega$.}
    \label{N_and_O}    
\end{figure}

Through our grid search, we immediately find that independent of $K$, increasing the bandwidth $\Omega$ decreases both the MSE between $\psi(\tau)$ and $f(\tau)$ and the maximum amplitude of $\psi(t)$ (Fig. \ref{N_and_O}, top). Further, independent of $\Omega$, increasing $K$ leads to a smaller MSE but increases lobe sizes (Fig. \ref{N_and_O}, bottom), until a threshold for $K$ (denoted as $K^{*}$ from here on) is reached. After this threshold, MSE does not appreciably improve and $\psi_{max}$ no longer increases as $K$ increases. Near-perfect recreations of $\psi$ therefore occur for any $K \geq K^{*}$, at the expense of near-maximal lobe amplitudes. 

\begin{figure}[H]
    \centering
    \includegraphics[width=.95\textwidth]{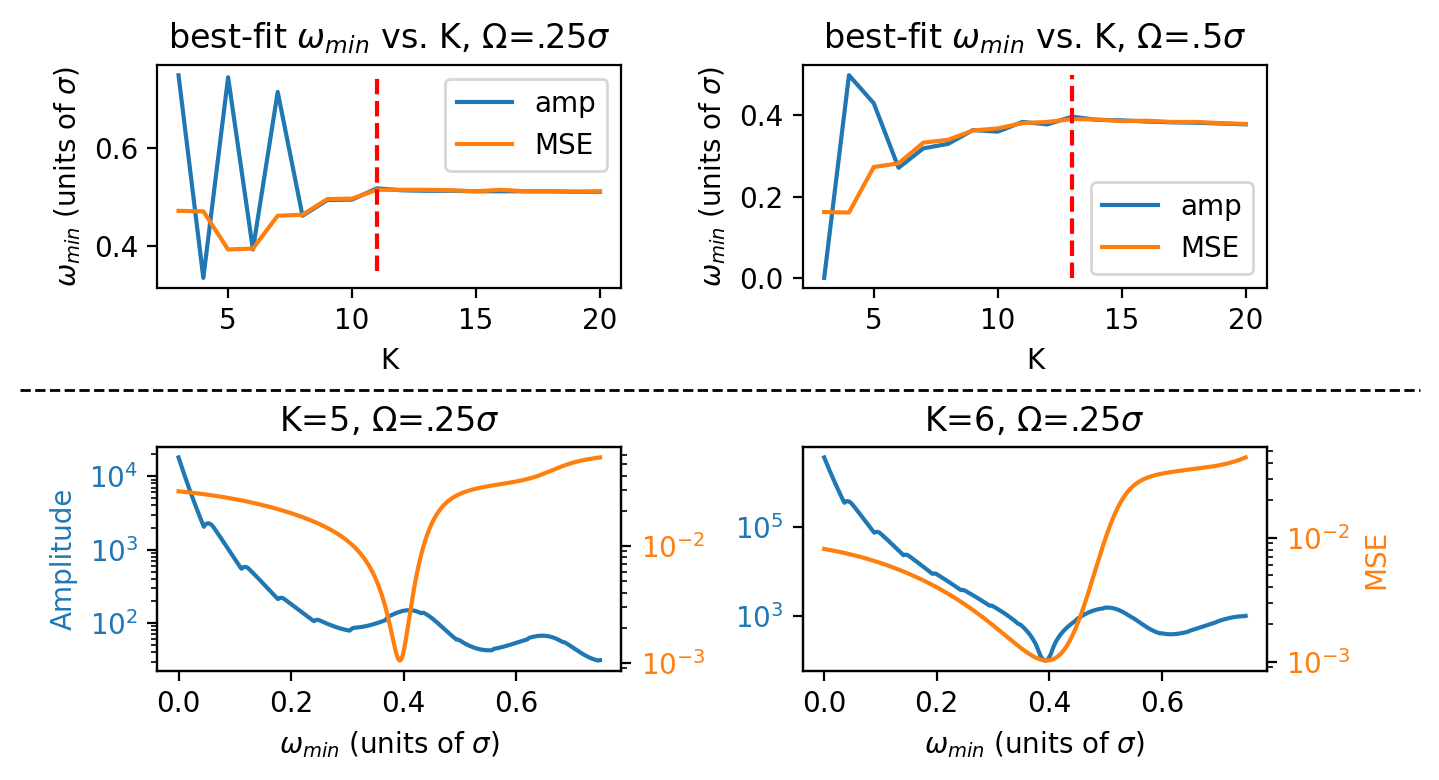}
    \caption{Top: plots of $\Omega=.25\sigma$ (top left) and $\Omega=.5\sigma$ (top right) show how the ideal $\omega_{min}$ trends toward the same value for both minimizing MSE and minimizing lobe amplitude. $\omega_{min}$ maximizes at $K=K^*$ in both cases ($K^*=11$ for $\Omega=.25\sigma$ and $K^*=13$ for $\Omega=.5\sigma$).
    Bottom: Plots of $K=5$ (bottom left) and $K=6$ (bottom right) show the impact of varying $\omega_{min}$ at a fixed $\Omega=.25\sigma$. While $K=5$ and $K=6$ have nearly identical minimum MSE values and corresponding $\omega_{min}$ locations, the amplitude values for $K=5$ are higher at this point, making $K=5$ an unideal candidate for minimizing lobe amplitudes.}
    \label{2}    
\end{figure}

Perhaps more interestingly, we find that for any $\Omega$, as $K$ increases the $\omega_{min}$ value that minimizes the MSE between the target sinc function and its frequency-comb approximation is \textit{also} the $\omega_{min}$ value that minimizes the amplitude of the lobes (Fig. \ref{2}, top). There is therefore always a singular strictly-best $\omega_{min}$ value for any ($\Omega, K$) combination with sufficiently large $K$. $K^*$ functions as a limit here too: the $\omega_{min}$ of best fit is at its maximal value when $K=K^*$. For smaller $K$ values, reconstructions seem to be too unstable to consistently find $\omega_{min}$ values that minimize both lobe amplitude and MSE. This leads to, for example, larger amplitudes for $K=5$'s best MSE fit than for $K=6$'s best fit. (Fig. \ref{2}, bottom). 

For $K < 5$, the quality of sinc reconstruction is too poor to use (as can be seen by their large MSE values relative to those of larger $K$ values in Fig. \ref{N_and_O}, bottom; quantitatively, any wave construction that resulted in an MSE larger than .01 was not considered). $K=6$ was therefore chosen to achieve the smallest possible lobe amplitude values while still maintaining a high degree of fitting accuracy. Additionally, $\Omega=.25\sigma$ was chosen arbitrarily to demonstrate a superoscillating wave's ability to mimic a function with four times the bandwidth while still keeping reasonable lobe sizes.

Lastly, $\omega_{min}=.4\sigma$ was chosen primarily for its proximity to $.394\sigma$, which is the $\omega_{min}$ value that minimizes both $\psi_{max}$ and $MSE$ for $K=6,$ $\Omega=.25$. This value was rounded to .4 to make each $\omega_{k \in K}$ an integer multiple of $\Delta\omega = \abs{\omega_{1} - \omega_{0}} = \frac{\Omega}{K-1} = .05\sigma$. This made spectral filtering at specific frequencies dramatically easier and more accurate. This can be easily understood by first knowing that, assuming evenly-spaced frequencies, one period of a frequency comb wave is the inverse of the spacing between teeth,
\begin{equation}
    T = \frac{1}{\Delta\omega}.
    \label{T}
\end{equation}
Fourier transforming one complete period $T$ of $\psi$ will therefore use $\Delta\omega$ as its frequency sampling rate ($\Delta f$), with all frequencies sampled as integer multiples of $\Delta\omega$. Further, any sampled integer multiple $r$ of $T$ for $\psi(t \in rT)$ will result in frequency sampling rates $\Delta f$ that are fractions of $\Delta\omega$ by $r$, 
\begin{equation}
    r \Delta f = \Delta\omega. 
    \label{dOmega}
\end{equation}
It follows, then, that any integer multiple of $\Delta\omega$ will also be an integer multiple of $\Delta f$. This assures that, by setting all $\omega_{k \in K}$ to be integer multiples of $\Delta\omega$, the Fourier transform of any integer $r$ periods of $\psi(t)$ will always result in a sampling rate that perfectly samples each frequency component of $\psi$ at its exact frequency value. This is crucial for the success of extreme spectral filtering by discrete frequencies.

\end{document}